\documentclass[12pt]{article}
\usepackage{graphicx}
\title{\bf Bringing Science into Schools through Astronomy.
Project ASTRO, Tucson.
 }
\author{Caroline Barban$^{1}$ \& Herv\'e Dole$^2$
\vspace{0.4cm}\\
\normalsize $^1$ Instituut voor Sterrenkunde, Katholieke Universiteit
Leuven, Celestijnenlaan 200 B, \\
\normalsize 3001 Leuven, Belgium\\
\normalsize $^2$Institut d'Astrophysique Spatiale, b\^at. 121, Universit\'e Paris
Sud 11, 91405 Orsay, France}
\date{\mbox{}}
\begin{document}
\maketitle
\pagestyle{empty}
%
%
\def\bull{\vrule height .9ex width .8ex depth -.1ex}
\makeatletter
\def\ps@plain{\let\@mkboth\gobbletwo
\def\@oddhead{}\def\@oddfoot{\hfil\tiny\bull\quad
``Science Case for Next Generation Optical/Infrared Interferometric Facilities 
(the post VLTI era)'';
37$^{\mbox{\rm th}}$ Li\`ege\ Int.\ Astroph.\ Coll., 2004\quad\bull}%
\def\@evenhead{}\let\@evenfoot\@oddfoot}
\makeatother
%
%
\def\beginrefer{\section*{References}%
\begin{quotation}\mbox{}\par}
\def\refer#1\par{{\setlength{\parindent}{-\leftmargin}\indent#1\par}}
\def\endrefer{\end{quotation}}
%
%
%
{\noindent\small{\bf Abstract:} We report our experience in bringing
science into US and French classrooms.  We participated in the US
scientific educational program Project ASTRO. It is based on a
partnership between a school teacher and an astronomer. They together
design and realize simple and interesting scientific activities for
the children to learn and enjoy science.  We present four hands-on
activities we realized in a 4th-grade class (10 yr-old kids) in Tucson
(USA) in 2002-2003. Among the covered topics were: the Solar System,
the Sun (helioseismology) and the Galaxies.  We also present a similar
experience done in two classrooms in 2005, in Ch\^atenay-Malabry
(France) in partnership with an amateur astronomy association
(Aph\'elie), and discuss future activities. This is a pleasant and
rewarding activity, extremely well appreciated by the children and the
school teachers. It furthermore promotes already at a young age the
excitement of science, and provides concrete examples of the
scientific methodology.}

%
%
%
\section{Introduction}
Many astrophysicists worldwide have experienced the joy of talking
about the physical Universe and their fascinating work to the public,
in conferences or in schools. They might have then felt the enthusiasm
and the great demand from children, their teachers and parents to talk
about science. The scientific education, an important cornerstone in
modern societies, faces, nevertheless, several difficulties in
practice despite national requirements for science classes. The
difficulties to teach science in schools (for students younger than
17) are sometimes related to the teachers' lack of confidence in their
own expertise. Science might appear complex, too conceptual and not
enough concrete to some teachers and students. The lack of motivation
might also be an issue. There is thus a clear need to bring science in
the classroom, so that more teachers and students can enjoy it.  We
describe our involvement in different educational programs in the USA
and in France to promote science among teachers and students.

%
\section{Project ASTRO, Tucson}

Project
ASTRO\footnote{http://www.astrosociety.org/education/astro/project\_astro.html}
was developed in 1994 by the Astronomical Society of the Pacific. The
goal of this program is to improve science education in the classroom
from the first year of elementary school to the last year of high
school (6 to 17 yr-old).  Project ASTRO is based on a partnership
between a teacher and an astronomer. The astronomer commits to
visiting the teacher's classroom at least four times during the school
year. The main focus of Project ASTRO is hands-on activities related
to astronomy that put students in the position of acting like
scientists. Each year about 1000 partnerships bring science through
astronomy to $\sim$100,000 students over 12 project ASTRO sites across
the USA.


We were part of Project ASTRO in
Tucson\footnote{http://www.noao.edu/education/astro/ It is managed in
Tucson by the Education and Outreach department of the NOAO (National
Optical Astronomy Observatory).} during the 2002-2003 school year. We
were paired with a 4th grade teacher (10 yr-old kids) at the Sewell
Elementary School in Tucson (Arizona, USA). The topics of the
activities were chosen in consultation with the teacher.  We designed
ourselves these activities, but extensively used existing material
from the
book\footnote{http://www.astrosociety.org/education/astro/astropubs/astropubs.html}
``The Universe at your fingertips'', and were inspired by some NASA
Education web sites.  We decided to focus on a few simple concepts
only in each one-hour visit. The typical visit starts with an
introductory 10-minute slide show, and is followed by 40 minutes of
hands-on activity. The last 10 minutes are used for a debriefing and
questions. We put
online\footnote{http://lully.as.arizona.edu/$\sim$hdole/vulgarisation/ProjectAstro200203/}
all our material.

%
\vspace{0.3cm}
\noindent $\bullet$ {\bf First visit: "Who is an astronomer"}\\
Our first visit to the classroom is dedicated to talk about who an
astronomer is and what an astronomer does. We first ask the children to
draw what they think an astronomer is, and then we use the drawings to
address the misconceptions they might have about this special job.  It
is a good opportunity to mention that an astronomer, thus a scientist,
is not always an old man! We also tell the students that an
astronomer does not work every night by looking the sky with the eyes
on a telescope. We introduce them to the concept that observation,
careful analysis, modeling, theory and simulation complement each other
and are required to address a scientific question.

  \begin{figure}
  \begin{minipage}{8cm}
  \centering
  \includegraphics[width=7cm]{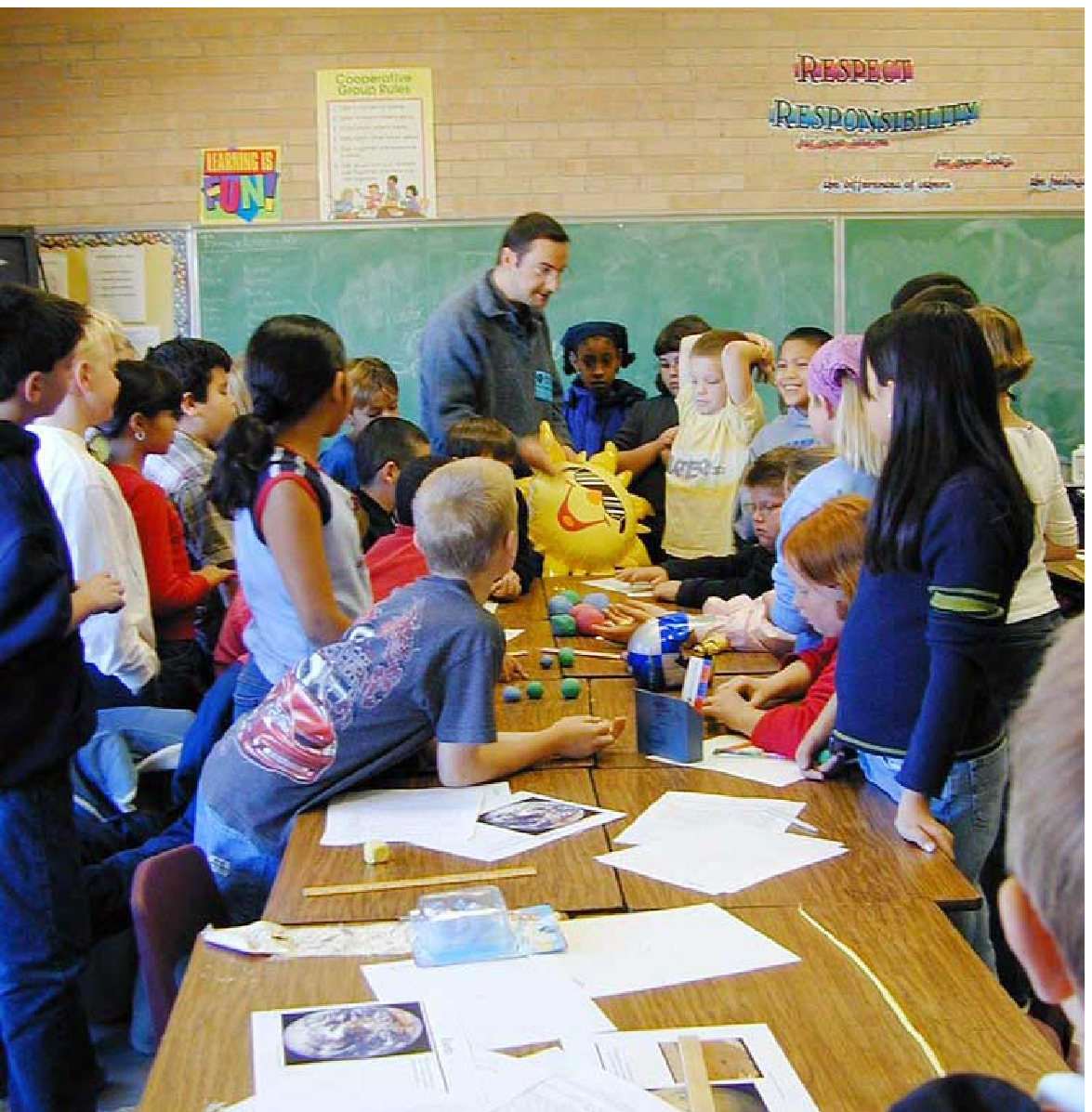}
  \caption{Solar System activity: comparison of the Planet's sizes.}\label{planets}
  \end{minipage}
  \hfill
  \begin{minipage}{8cm}
  \centering
  \includegraphics[width=5.3cm]{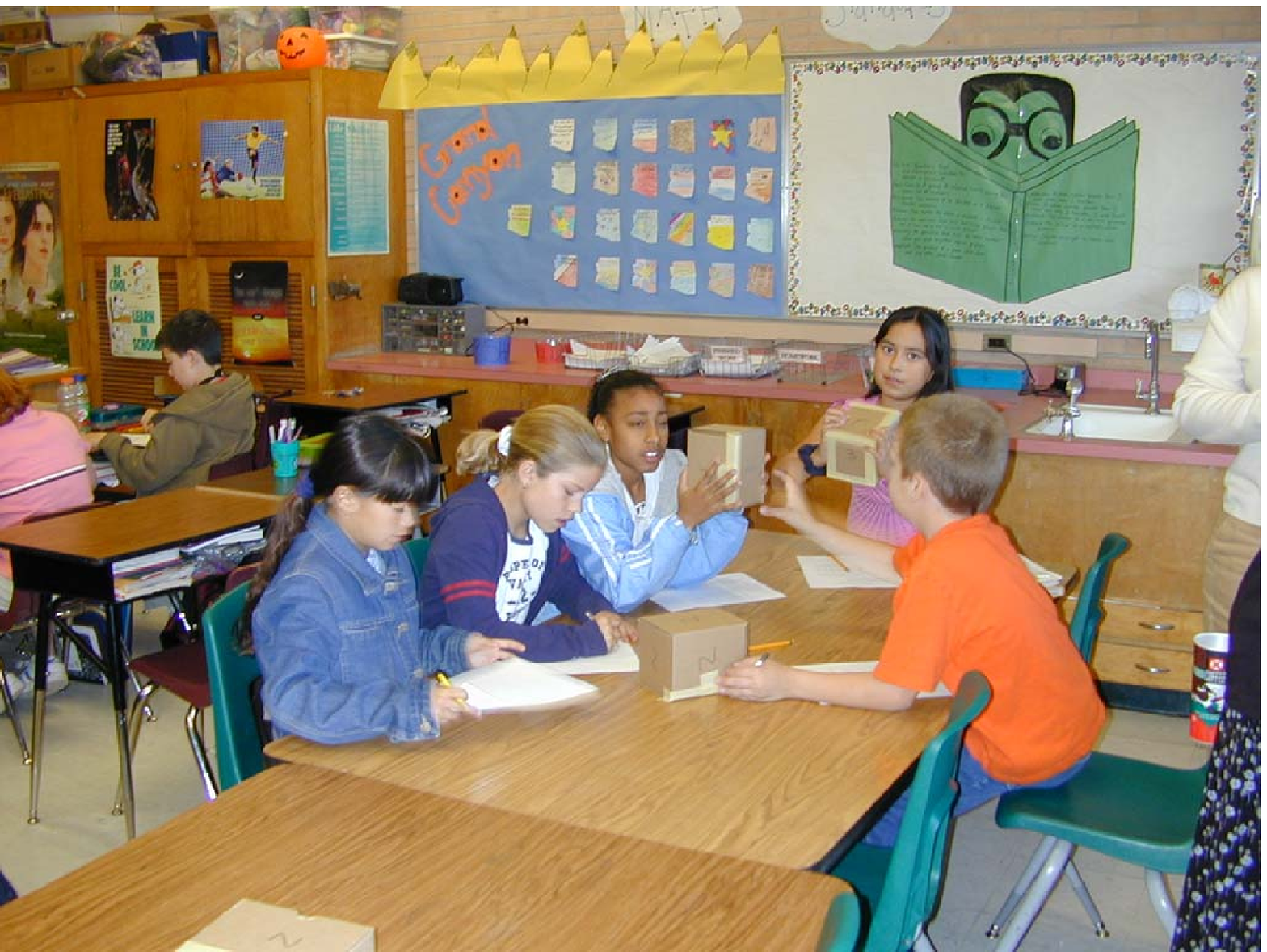}
  \caption{Helioseismology activity: probing the mystery boxes.}\label{sun}
  \includegraphics[width=5.3cm]{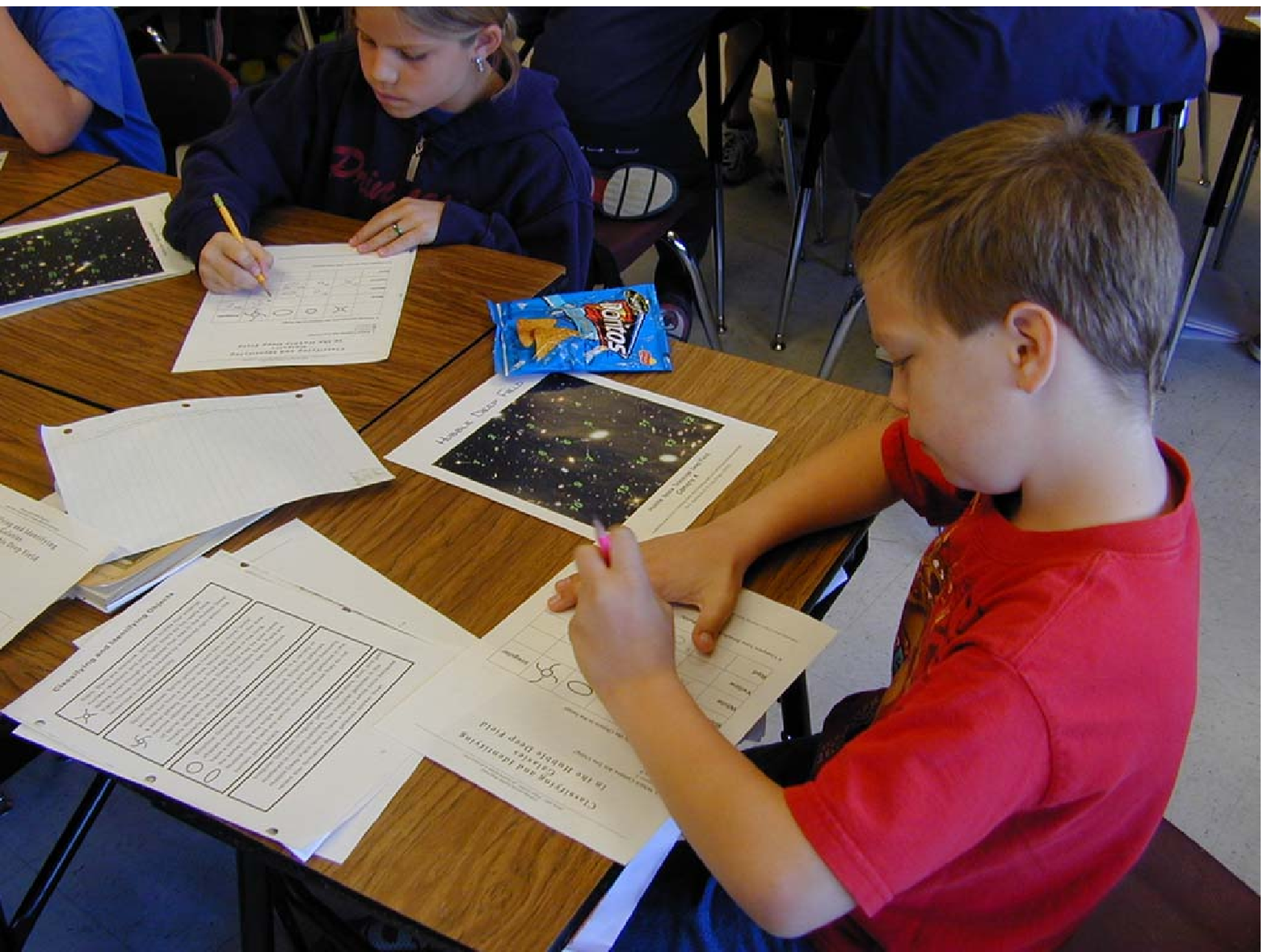}
  \caption{Galaxy activity: classifying the galaxies in the
   Hubble Deep Field.}\label{galaxies}
  \end{minipage}
  \end{figure}

%
\vspace{0.3cm}
\noindent $\bullet$ {\bf Second visit: "The Solar System"}\\
The goal of this activity is to learn the different planets of our
Solar System and the relative sizes of the planets.  Each kid makes a
sphere of the proper diameter with clay. We then put together this
clay Solar System, describe each Planet's average temperature and
emphasize the diversity of planet sizes (Fig.~\ref{planets}). We also
explore the physical differences between gaseous vs telluric planets,
and discuss the differences between the Planets, the Sun and the
stars.

%
\vspace{0.3cm}
\noindent $\bullet$ {\bf Third visit: "The Music of the Sun"}\\
This activity, dealing with the Sun and helioseismology, illustrates
that we can get informations about the interior of an object by
"listening" to it. The students have to probe three ``mystery
boxes''. We ask them to shake each box to study its content. They
estimate if the content is light or heavy, composed of small or large
pieces, in small or large number. They write their observations in a
table. We put rice, small rocks, and a tennis ball in each box,
respectively. We then explain that helioseismology uses the Sun's
vibrations (like music) to probe its interior.  This activity shows
that it is possible to talk about complex subjects in a simple way
(Fig.~\ref{sun}). The last part is an illustration of the Doppler
shift (also used in helioseismology): listening to the sound emitted
by an object allows one to determine if it is moving or not. We use a
bell for this purpose. We first introduce them to the bell's sound
when it is still, but also when it is rapidly spinning -- the beating
effect is then easily heard. Then the students close their eyes, and we
ask them to guess, while listening the sound, if the bell is spinning
or not.

%
\vspace{0.3cm}
\noindent $\bullet$ {\bf Fourth visit: "The Galaxies"}\\
The goal of this activity is to have the children realize the
diversity of the galaxies and their incredible quantity. Using images
of the famous Hubble Deep Field, the students sort the bright sources
in the images by shape and color. They separate foreground stars from
far-away galaxies, and get to realize that the galaxies have various
shapes and colors, mainly due to their different content and evolution
stage (Fig.~\ref{galaxies}). Then, on a small image cell, they have to
count the galaxies. With the appropriate provided number to multiply
with, they have an estimate (lower limit) of the number of visible
galaxies in the universe. Classifying the galaxies was a success, but
the counting part proved to be too complicated for this level, since
orders of magnitude and powers of ten are not well understood.

%
\section{Our Other Experiences in France} 
\begin{figure}
  \begin{minipage}{8cm}
  \centering
  \includegraphics[width=7cm]{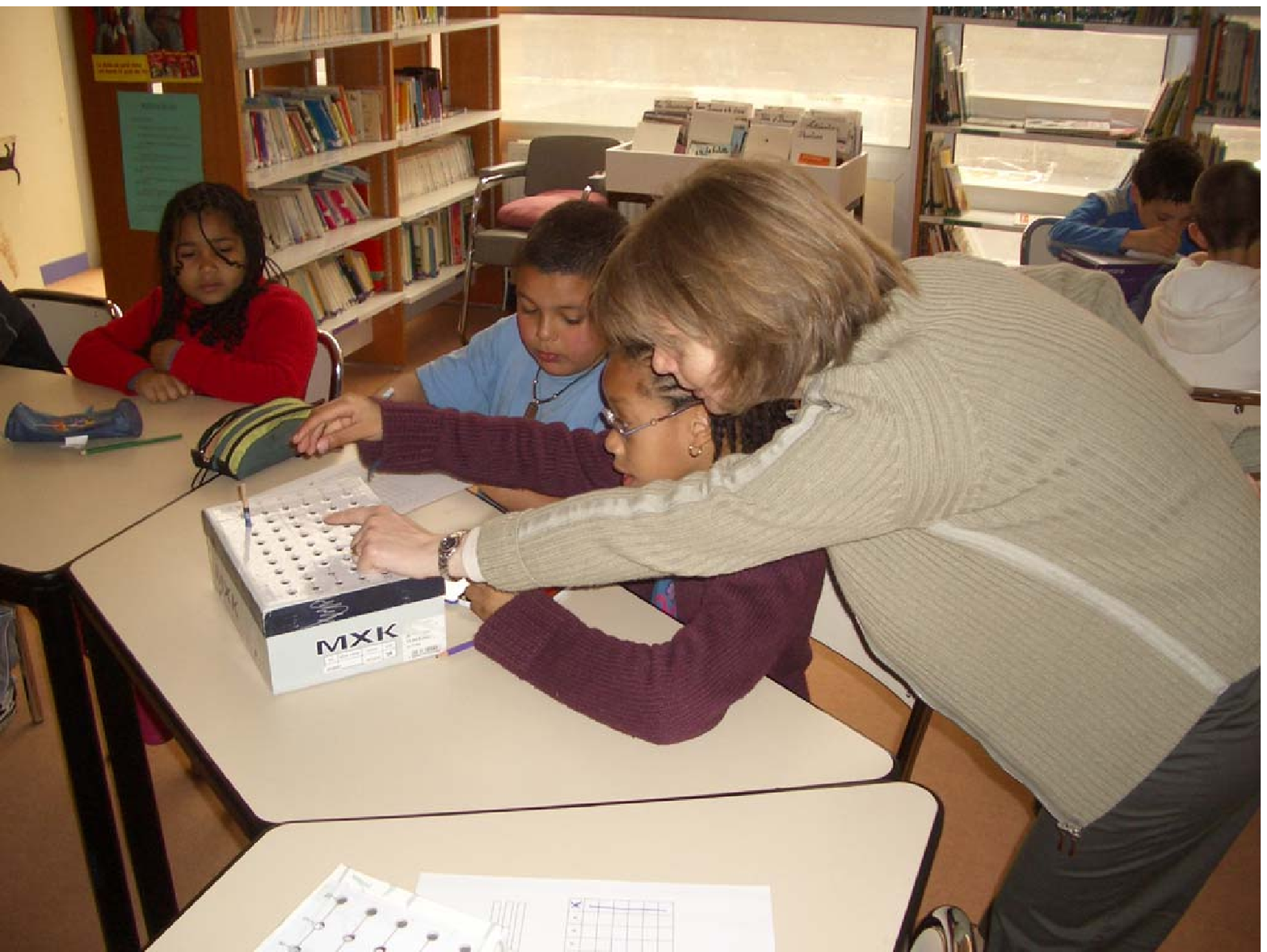}
  \caption{Venus topography box activity: probing the hidden surface.}\label{venus1}
  \end{minipage}
  \hfill
  \begin{minipage}{8cm}
  \centering
  \includegraphics[width=7cm]{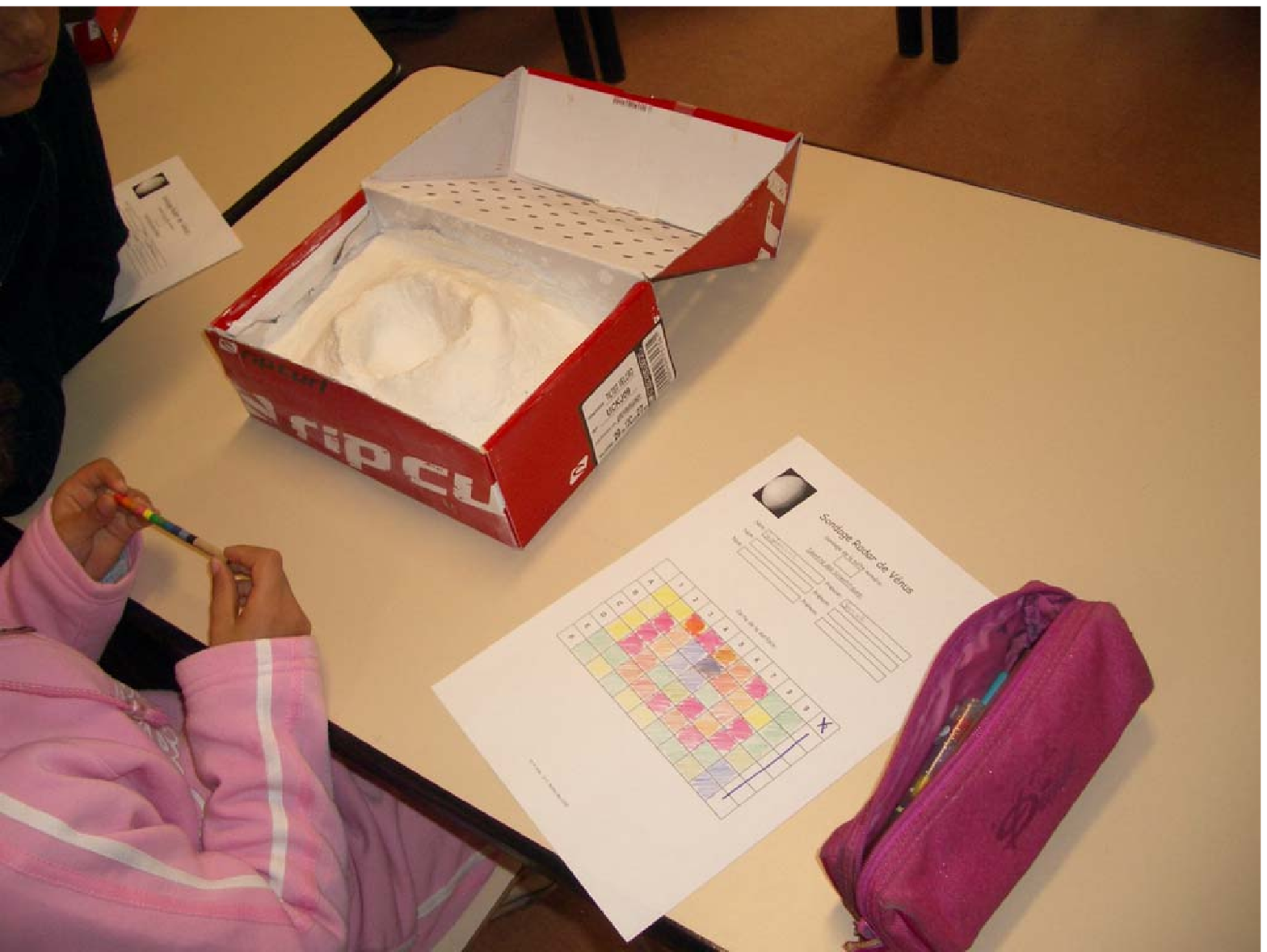}
  \caption{Venus topography box activity: at the end, comparing the measurements
(color map) and the real Planet surface model.}\label{venus2}
  \end{minipage}
  \end{figure}

There are many other ways to bring science into schools. As members of
the amateur astronomy association
Aph\'elie\footnote{http://www.astrosurf.com/aphelie} (near Paris,
France), we are used to receiving lots of requests from schools to
come and talk about astronomy. In May 2005, we visited two
classrooms\footnote{http://lully.as.arizona.edu/$\sim$hdole/vulgarisation/ecoles/}
in two of Ch\^atenay-Malabry's schools located in socially disfavored
areas. The level is CE2 and CM2 (8 and 10 yr-old, respectively). We
wanted to apply the concept of a Project ASTRO visit. We choose to try
``{\bf the Venus topography box}'' activity from the Project ASTRO
book$^3$, in agreement with the teachers.  The goal is to show how
scientists get information about the surface of a planet hidden by a
thick atmosphere (this could also apply to Titan). After an overview
of the Solar System, we tell the children that ESA wants to send a
spacecraft to Venus and needs their help to choose the best landing
site. The surface is modeled inside a shoe box, hidden by a cover. By
probing the holes with a color-encoded stick, the students draw a
topographic map of the modeled Venus surface (Fig.~\ref{venus1} \&
\ref{venus2}). Fig.~\ref{venus2} shows the model crater (top left)
and the beautiful result, the crater appearing in pink-orange in the
topographic map (lower right). We then asked each student to present
in front of the class their best landing site, if any. This activity
is very interesting because it deals with different concepts: variety
in the Solar System, thick atmospheres, remote information gathering
(e.g radar technique), contour map. During this activity the students
really act like scientists by getting the data, and then analyzing,
interpreting and presenting them.

%
\section{Conclusion}

Some professional astrophysical institutions are also involved in
bringing science into the classrooms through specific educational
programs for the teachers, like the Observatoire de
Paris\footnote{http://www.obspm.fr/$\sim$webufe} (dedicated workshops,
summer schools, observations, ...) or the Universit\'e Paris Sud
11. There are also many successful programs like ``la main \`a la
p\^ate\footnote{http://www.lamap.fr/ --
http://www.handsonuniverse.org/ and http://fhou.cicrp.jussieu.fr/ --
http://www.ac-nice.fr/clea/}'', Hands-On-Universe$^8$, or the
CLEA$^8$, among others.

Finally, we want to emphasize that visiting schools, discussing with
the teachers, interacting with the students, and designing such
educational projects is a rewarding, interesting and very useful
activity. Astronomy is a very efficient way to teach
science. Furthermore this is less time-consuming that one might
think. We warmly recommend that our colleagues try to be involved in
such activities, on their own or by joining an existing program in
their country.

%
%

\vfill  
\vfill

%
%
%
\vspace{-0.5cm}
\section*{Acknowledgments}
\vspace{-0.3cm}
{\small Thanks to Connie Walker (NOAO), Project ASTRO (ASP), Pam
Williams (Sewell Elementary), GONG/National Solar Observatory,
MIPS/Steward Observatory/University of Arizona. We also warmly thank
Marc Bottineau (for the 15 boxes !) and the Aph\'elie members involved in
this activity, Fabrice Krot and the Maison des Sciences de
Ch\^atenay-Malabry, as well as S\'ebastien Vilain-Derouen (\'ecole
\'el\'ementaire T. Masaryk) and Claude Mallet (\'ecole \'el\'ementaire
L. De Vinci).}

%
%
 

\end{document}